\documentstyle[twoside,fleqn,espcrc2,graphicx]{article}

\newcommand{\AmS}{{\protect\the\textfont2
  A\kern-.1667em\lower.5ex\hbox{M}\kern-.125emS}}

\title{Observation of coherent Josephson response in the non-linear ab-plane microwave impedance of $YBa_{2}Cu_{3}O_{6.95}$ single crystals}

\author{Z.Zhai, H.Srikanth and S.Sridhar, \address{Physics Department, Northeastern University, 360 Huntington Ave, Boston, MA 02115} 
        \thanks{Work supported by NSF-DMR-9623720}
        and 
        A.Erb, E.Walker and R.Flukiger \address{DPMC, Universite de Geneve, CH-1211 Geneve, Switzerland}}

\begin{document}

\begin{abstract}
We report novel non-linear phenomena in the $ab$-plane microwave impedance
of $YBaCu_{2}O_{7-\delta }$ single crystals. The $R_s$ vs. $H_{rf}$ data are well described by
the non-linear RSJ model : $\dot{\phi}+\sin \phi =i_{rf}\cos \omega t$. The
entire crystal behaves like a single Josephson junction. The extraordinary
coherence of the data suggests an intrinsic mechanism.

\end{abstract}
\maketitle

We describe some striking observations associated with the microwave
ab-plane response of high quality $YBaCu_{2}O_{7-\delta }$ crystals\cite{AErb96a}. These result from our accidental observation that the microwave surface impedance $%
Z_{s}=R_{s}+iX_{s}$ of crystals is {\em non-linear at very low applied
microwave fields }$H_{rf}$. The analysis\cite{TJacobs95e} of the results
implies that the entire crystal responds like a single Josephson junction to
an ac drive current in the ab-plane.

\begin{figure}[htb]
\begin{center}
  \includegraphics*[width=0.45\textwidth]{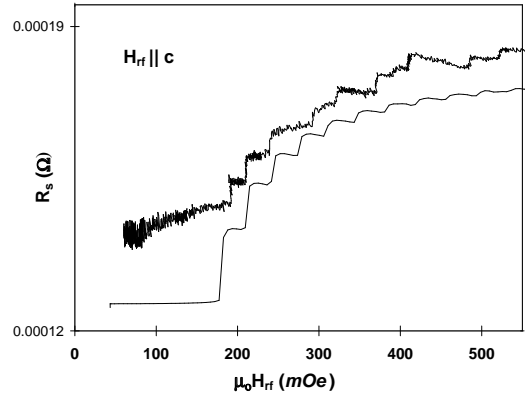} 
\end{center}
\caption{Experimental data for $R_s$ vs. $H_{rf}$ at $T = 4.2K$for a YBCO crystal, and results (solid line) of numerical calculations using the
non-linear RSJ model. The model results are displaced below the data for clarity.}
\label{Fig1}
\end{figure} 

Microwave experiments were carried out in a Nb cavity at 10 GHz with the microwave field $H_{rf} \parallel c$\cite{DHWu90a}. The $R_{s}$ vs. $H_{rf}$ $(\parallel c)$data in Fig.~\ref{Fig1} shows a
region of linear response at extremely low fields in which $R_{s}$ is
independent of $H_{rf}$. The $T$- dependence of these high quality crystals
is discussed in a separate publication\cite{Srikanth96a}. As the microwave
field $H_{rf}$ is increased, an onset of increased absorption is evident at
a threshold value we call $H_{crit}$, which is about $200mOe$ at low $T$ and
decreases with increasing $T$. For further increase of $H_{rf}$ there is an
overall parabolic rise of $R_{s}$, however there are definite jumps steps followed by plateaus at several values of the increasing $H_{rf}$. It should be noted that even at the highest field the overall change is very small and
represents only a $50\%$ increase of $R_{s}$. At $4.2K$, this overall change
is still {\em smaller by }$10^{3}$ {\em than the normal state value }$R_{n}$.

Remarkably, a very simple model completely describes the essential features
of the $R_{s}$ and $X_{s}$ data. Consider a resistively-shunted Josephson
junction (RSJ). The junction phase $\phi $ obeys the dynamical ``overdamped
pendulum'' equation $\beta ^{-1}\dot{\phi}+\sin \phi =(i_{rf}/I_{c})\cos \omega t$
where we consider a {\em pure ac drive }$i_{rf}\cos \omega t$, and $\beta
=2eI_{c}R/{\hbar}$. The RSJ equation is better analyzed in dimensionless form $%
d\phi /d\tau +\sin \phi =(i_{rf}/I_{c})\cos \Omega \tau $, where $\tau =\beta t$,
and $\Omega =\omega /\beta $. Unlike most junction measurements our
experiments measure the dynamic ac impedance and not the dc I-V
characteristics. The ac impedance can be calculated \cite{Sridhar96,Xie96,McDonald96} from the Fourier
components at $\omega $ of the voltage $\dot{\phi},$ $Z_{\omega }=R_{\omega
}+iX_{\omega }=(\omega /2\pi i_{rf})\int_{0}^{2\pi /\omega }\dot{\phi}%
e^{i\omega t}dt\,$. The result of numerical calculations for $R_{\omega }$
are also shown in Fig.~\ref{Fig1} for $\Omega =\omega /\beta =0.08$. All the
features of the $R_{s}$ vs. $H_{rf}$ data are very well reproduced - the
threshold at $i_{rf}\approx 1$, the parabolic rise and the occurrence of
steps and plateaus for $i_{rf}>1$.

From $\omega =2\pi 10^{10}$ $rad/sec$ and $\Omega =0.08$, we deduce an $%
I_{c}R$ product of $I_{c}R=0.3mV$, remarkably close to those reported in the
literature for fabricated Josephson junctions of YBCO. This value
also implies that inertial effects associated with a $\ddot{\phi}$ term can
only be observed at frequencies $\omega >\beta ^{-1}\approx 125GHz$.

The threshold value observed in Fig.~\ref{Fig1} is $H_{crit}\sim 200mOe$ at $%
4.2K$. This is an extremely low value compared to estimates of $H_{c1}\sim
250\,Oe$ \cite{DHWu90a}, even correcting by a factor of $10$ for
demagnetization, and hence cannot be associated with entry of Abrikosov
vortices into the sample. Additional experiments have confirmed that trapped vortices are not responsible for the observed effect.

The excellent agreement between the experimental data and the very simple
RSJ model strongly suggests that {\em the entire macroscopic crystal behaves
as a single Josephson junction}.

The presence of ``weak links'' is well known to be responsible for the
microwave response of ceramics but is less likely in crystals, even though they are twinned. However if defects are responsible, then they must respond in an extraordinarily coherent manner to account for the entire crystal responding like a single JJ, and this appears rather unlikely but cannot be ruled out.

The extraordinary coherence of the data in Fig.\ref{Fig1} suggests the possibility that we are observing an intrinsic effect. Recently we have analyzed the {\em linear response} penetration depth $\lambda (T)$ and conductivities $\sigma _{1}(T)$ and $\sigma _{2}(T)$ of the same crystal\cite{Srikanth96a}, and observed new effects in the temperature dependence also. The $T$-dependent data suggest that a simple picture of a single order parameter (whether of $d$-wave or any other symmetry) is not correct, and instead there are at least two superconducting channels in $YBaCu_{2}O_{7-\delta }$.

The presence of multiple superconducting channels raises the possibility of
coherent tunneling between them. The possibility of  an ``internal'' or ``bulk'' Josephson effect in a two-band or two-gap superconductor has been suggested before \cite{AJLeggett66,Rogovin71,Vittoria93}. In a multi-component superconductor an external current can induce a phase difference between the two order parameters and this could be responsible for the observed effects reported here. 

Our experiment further suggests that magnetic fields penetrate at very low field strengths well below $H_{c1}$ in the form of fluxons even in high quality crystals. When the applied field is oscillating this leads to phase slip which we are able to observe as steps in the non-linear impedance data. Current relaxation at high frequencies takes place by this phase slip process and is responsible for the electrodynamic response at low applied fields and high frequencies.

\end{document}